\documentclass[a4paper,prb,twocolumn,showpacs]{revtex4}

\usepackage{amsfonts}
\usepackage{graphicx}
\usepackage{color}
\usepackage{amsmath}
\usepackage{amssymb}
\usepackage{latexsym}
\usepackage{psfrag}

\begin{document}
\title{Activated hopping transport in anisotropic systems at low temperatures}

\author{S. Ihnatsenka}
\affiliation{Laboratory of Organic Electronics, ITN, Link\"{o}ping University, SE-60174
Norrk\"{o}ping, Sweden}

\email{siarhei.ihnatsenka@gmail.com}

\begin{abstract}
Numerical calculations of anisotropic hopping transport based on the resistor network model are presented. Conductivity is shown to follow the stretched exponential dependence on temperature with exponents increasing from 1/4 to 1 as the wave functions become anisotropic and their localization length in the direction of charge transport decreases. For sufficiently strong anisotropy, this results in nearest-neighbor hopping at low temperatures due to the formation of a single conduction path, which adjusts in the planes where the wave functions overlap strongly. In the perpendicular direction, charge transport follows variable-range hopping, a behavior that agrees with experimental data on organic semiconductors.
\end{abstract}

\maketitle

\section{Introduction}
Electrical conduction in organic semiconductors is typically interpreted in terms of temperature activated hopping of charge carriers. A seminal work\cite{Mot69} by Mott showed that the hopping conductivity follows a stretched exponential dependence on temperature,
\begin{align} \label{eq:1}
 \sigma&=\sigma_0\exp\left[-\left(\frac{T_0}{T}\right)^{\alpha}\right],\\
 T_0&=\frac{\beta}{\rho_0\xi^d},
\end{align}
where $\alpha=1/(1+d)$, $d$ is dimensionality, $\rho_0$ is the density of localized states at the Fermi level, $\xi$ is the isotropic localization length proportional to the carrier wave function extent and $\beta$ is a numerical coefficient ($\beta=21.2$ and 13.8 for $d=3$ and 2, respectively\cite{Shk_book}). In the derivation of \eqref{eq:1}, the charge transport was assumed to be dominated by the states within a narrow energy band close to the Fermi energy, and within that energy band the charge transport occurs by variable-range hopping (VRH).\cite{Shk_book} It becomes nearest-neighbor hopping (NNH) for $\alpha=1$ with $k_BT_0$ being the activated energy. Eq. \eqref{eq:1} has been routinely used to determine dimensionality,\cite{Kim12, Nar07_AM, Ash05, Ale04} $T_0$ and consequently $\xi$ if $\rho_0$ is known, or vice versa, from a separate measurement.\cite{Nar07_PRB, Ree99, Nar08, Shu12} Conductivity in a system having structural anisotropy is still expected to follow \eqref{eq:1} with the same $\alpha$ for all directions, but different $\sigma_0$, which becomes direction dependent and related to carrier wave function anisotropy. \cite{Shk_book}However, surprisingly, in experiments by Nardes \textit{et al.}\cite{Nar07_PRB}, thin films of poly(3,4-ethylenedioxythiophene) (PEDOT), which were prepared by spin coating, showed $\alpha=0.25$ for $\sigma$ measured in the lateral direction ($\sigma_{\parallel}$) and $\alpha=0.81$ for measurement in the perpendicular (vertical) direction ($\sigma_{\bot}$), with a ratio $\sigma_{\parallel}/\sigma_{\bot}=10-10^3$. This has led to a conclusion about VRH in the lateral and NNH in the vertical direction, but the microscopic origin of the co-existence of those two regimes remained an open question. Another uncertainty exists regarding the fractional value $\alpha=0.81$ that is less than 1 expected for activated Arrhenius-like transport. Fractional values of $\alpha$, which do not fit integral $d$, are commonly observed\cite{Kim12, Ale04} in conductivity measurements on organic semiconductors, which further lead to uncertainties in interpreting the morphology and nature of charge transport. 

The extraction of Mott's exponent $\alpha$ from the temperature dependence of conductivity is known to be error prone. The values extracted deviate commonly from 1/4, 1/3 and 1/2 that are characteristic to 3D, 2D and 1D charge transport, respectively. This led to conclusions of quasi-dimensional transport with morphology having no preferred dimensionality.\cite{Kim12} For $\alpha>1/2$, it was concluded about the transition between VRH and NNH.\cite{Nar07_PRB} A common method to obtain $\alpha$ is to plot $\sigma$ vs $T^{-\alpha}$ for different $\alpha$ and check whether it falls onto a straight line. The linearity could be then quantified via the correlation coefficient.\cite{Nar07_PRB, Nar08} Another, more accurate method is based on computing the reduced activation energy $d\log(\sigma)/d\log(T)$, for which a slope, when plotted as a function of $\log(T)$, directly gives $\alpha$.\cite{Zab84}

In this paper, numerical calculations of charge hopping transport in anisotropic systems are presented with a focus on an analysis of powers $\alpha$ entering the Mott's law \eqref{eq:1}. As the localized states become progressively anisotropic, $\sigma$ in a direction, where the localization length is smaller, follows \eqref{eq:1} with $\alpha$ taking any values between 1/4 and 1 at low $T$. This implies changing of VRH to NNH as a result of the formation of a single conduction path that carries most of the current. This is demonstrated by current visualization and also explained using the percolation theory. At the same time, $\sigma$ in a perpendicular direction retains VRH for any degree of anisotropy, which is all consistent with experimental data\cite{Nar07_AM, Nar07_PRB} on anisotropic conduction in PEDOT.

\section{Model}

The hopping conduction between localized states in a disordered system is modeled by a resistor network.\cite{Mil60, McI79, Amb73} The resistance between two states $i$ and $j$ is\cite{Shk_book}
\begin{equation} \label{eq:2}
 R_{ij}=\frac{k_BT}{e^2\Gamma_{ij}},
\end{equation}
where the average tunneling rate accounting for wave function anisotropy is
\begin{widetext}
\begin{equation} \label{eq:3}
\Gamma_{ij}=\gamma_0\exp\left(-2\sqrt{\frac{x_{ij}^2+z_{ij}^2}{\xi_{\parallel}^2}+\frac{y_{ij}^2}{\xi_{\bot}^2}} -\frac{|E_i-E_j|+|E_i|+|E_j|}{2k_BT}\right),
\end{equation}
\end{widetext}
with $\gamma_0$ being the electron-phonon coupling parameter, $\xi_{\parallel}$($\xi_{\bot}$) is the localization length in $xz$ plane ($y$ direction), see the inset in Fig. \ref{fig:1}(b), ($x_{ij}$, $y_{ij}$, $z_{ij}$) are coordinate components of the distance between states, and $E_i$ is the energy of the $i$-th state. The exponentially decaying wave functions are characterized by ellipsoids with semi-major and semi-minor axises $\xi_{\parallel}$ and $\xi_{\bot}$ (see inset in Fig. \ref{fig:1}(b)) that are centered on lattice sites of the cubic crystal that is assumed in the following. In this way $\xi_{\parallel}/\xi_{\bot}$ describes the degree of anisotropy; for the isotropic case $\xi_{\parallel}=\xi_{\bot}=\xi$ and \eqref{eq:3} reduces to a familiar expression for the tunneling rate.\cite{Shk_book} 
The linear Ohmic regime is assumed in the following and the chemical potential is set to zero. 

Applying the Kirchhoff's law to the resistor network, the resistance between two arbitrary nodes can be calculated\cite{Wu82} from the determinants of the conductance matrix $G$,
\begin{equation} \label{eq:4}
R_{ij}=\frac{|G^{ij}|}{|G^j|},
\end{equation}
where $|G^j|$ is the determinant of $G$ with the $j$-th row and column removed, and $|G^{ij}|$ is the same determinant but with the $i$-th and the $j$-th rows and columns removed. It is convenient to introduce two additional nodes serving as the source ($s$) and drain ($d$) electrodes and then connecting them to all nodes in the outer planes of the lattice by small resistances. Those nodes are substituted into \eqref{eq:4}, which is further used to compute conductivity,
\begin{equation}
  \sigma=\frac{1}{R_{sd}Nl},
\end{equation}
where $N$ is the edge length and $l$ is the constant of a cubic lattice. This method allows one to account for resistances between all pairs of the nodes in the system and thus current branching without any cut-off, which is more accurate than commonly implemented methods\cite{Amb73} and also the critical subnetwork approximation\cite{Amb71} used in the percolation approach.\cite{OPA} Note that a weak $T$ dependence due to the pre-exponential factor in \eqref{eq:2} is explicitly taken into account. To visualize the currents, the system of equations $I=GV$ is solved for a small source-to-drain voltage, $eV_{sd}\ll k_BT$.\cite{Amb73}

In the following, the $y$ axis is assumed to be a direction in which the anisotropic localized states are squeezed (Fig. \ref{fig:1}(b)), and if the source and drain electrodes align with the $y$ axis, it is said to be out-of-plane transport. If the electrodes are in the $x$ (or $z$) direction, transport is denoted as in-plane. 

\section{Results and discussion}

To analyze the influence of structural anisotropy on charge transport the numerical calculations are performed for a system with parameters typical for organic semiconductors.\cite{Kim12_Pipe} In particular, $\xi=\xi_{\parallel}$ is chosen to be equal to $l$, a value large enough not to bring the system into strong localization (insulating) regime. DOS is taken to be uniform (constant) with width $W$ (measured in units of Kelvin) that establishes an energy scale. The disorder is assumed to be only energetic; the effect of positional disorder will be later commented on. The system size for the results presented below is $20\times20\times20$. This allows to perform averaging over 10000 different disorder realizations within available computational resources. 
The calculations were also performed for different sizes and $\xi$ with similar results obtained.

\begin{figure}[h]
\includegraphics[keepaspectratio,width=0.9\columnwidth]{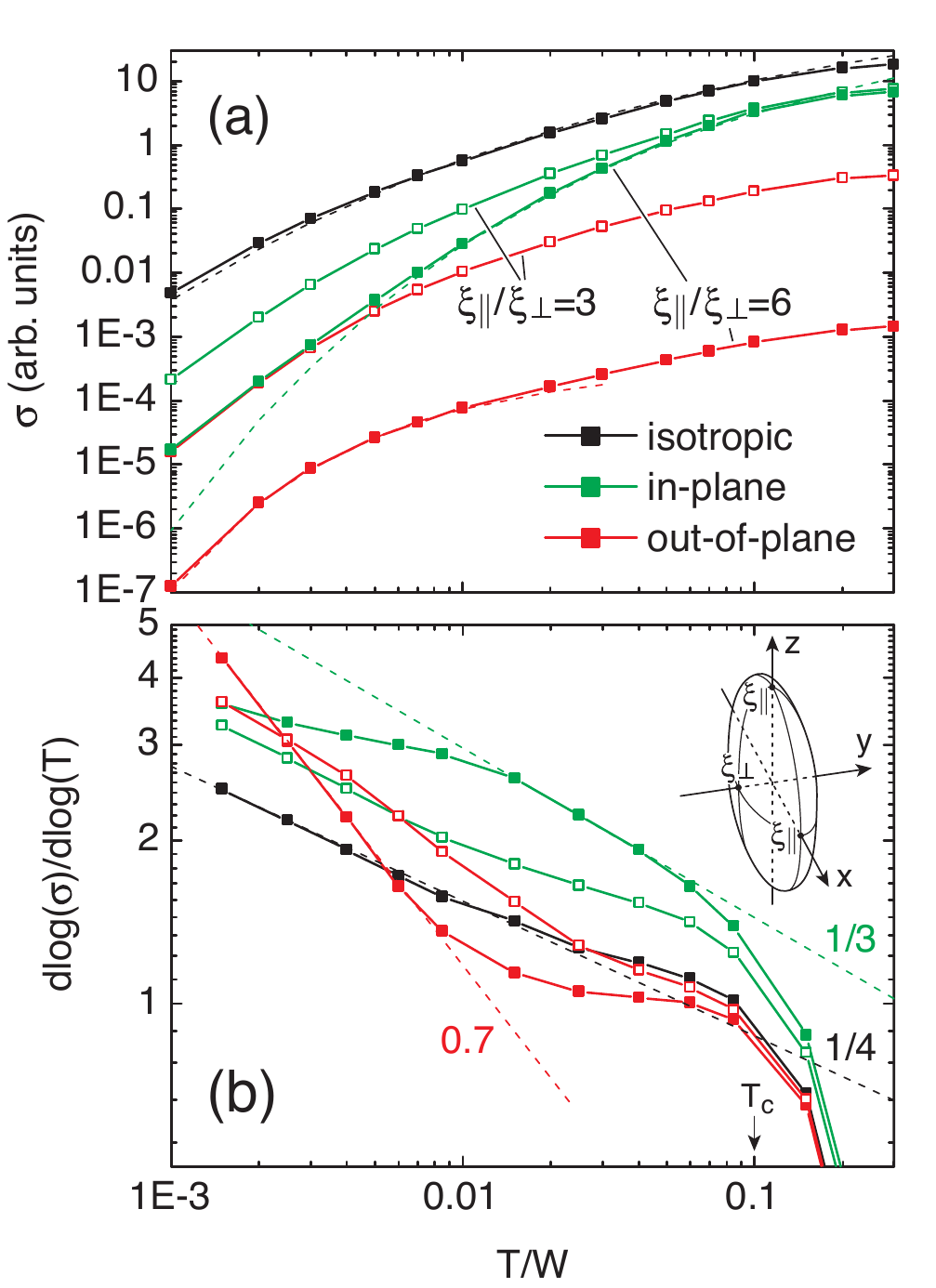}
\caption{(Color online) Temperature dependence of (a) averaged conductivity and (b) reduced activation energy. The dotted lines show a fit to Eq. \eqref{eq:1} with $\alpha$ denoted in (b). In an isotropic system, the localized states are spheres centered in the nodes of a cubic lattice, while the states in an anisotropic system are oblate spheroids squeezed in the $y$ direction as shown in the inset in (b). $\sigma$ in the $xz$ plane (in-plane), where neighbor states overlap more, and in the $y$ direction (out-of-plane) are shown for two values of anisotropy: $\xi_{\parallel}/\xi_{\bot}=3$ and 6. The lattice size is $20\times20\times20$.}
\label{fig:1}
\end{figure}

Figure 1 shows the temperature dependence of conductivity for different morphologies, as the localization states change from isotropic to anisotropic, for which the transport direction is either in-plane or out-of-plane. There, several transport regimes can be traced, which are easy to distinguish by slopes to $d\log(\sigma)/d\log(T)$ in Fig. \ref{fig:1}(b). At high temperatures ($T_c>0.1W$), conductivity follows activated behavior with $T_0/W\approx0.1$. This agrees with the traditional hopping theory\cite{Shk_book} that predicts activated transport for
\begin{equation} \label{eq:5}
 T_c>0.29W\rho_0^{1/3}\xi.
\end{equation}
At lower temperatures, VRH is observed with $\sigma$ described by the Mott's law \eqref{eq:1}. For the isotropic structure, $\alpha=1/4$ and $T_0/W=18$ are derived, while $\alpha=1/3$ and $T_0/W=7$ are derived for the in-plane conduction, implying $\beta=18$ and $\beta=7$ for 3D and 2D hopping, respectively, for $\xi=l$. These values agree well with known values\cite{Shk_book}, which, along with $T_c$ obtained above, justify the validity of the method implemented. \textit{While isotropic and in-plane hopping conduction demonstrate an expected behavior, out-of-plane conduction surprisingly reveals a reentrance to activated behavior at low $T$ as the anisotropy degree of the localized states becomes stronger.} For $\xi_{\parallel}/\xi_{\bot}=6$, $\alpha=0.7$, and it approaches 1 as the ratio $\xi_{\parallel}/\xi_{\bot}$ increases further.

\begin{figure*}[th!]
\includegraphics[keepaspectratio,width=\textwidth]{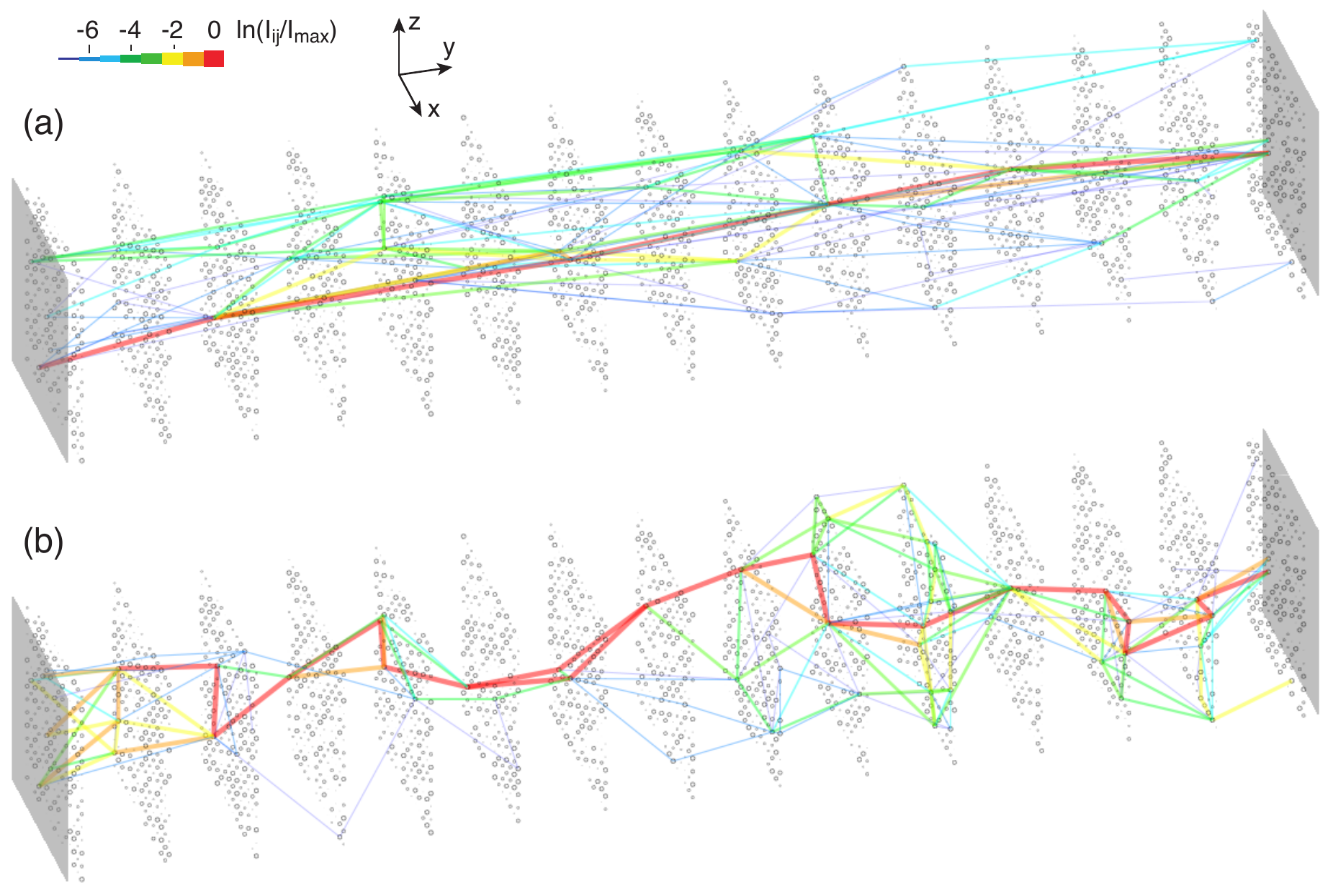}
\caption{(Color online) Currents in (a) isotropic and (b) anisotropic $\xi_{\parallel}/\xi_{\bot}=6$ structures mapped onto a view stretched along the $y$ axis: For better visualization, the distance between the $xz$ planes is intentionally increased after calculation has been done; the original lattice is cubic. The dots mark the hopping sites, with the dot size being inversely proportional to the absolute value of energy of the localized state. Gray pads are the source and drain electrodes. Both structures have a $15\times15\times15$ lattice size and an identical energetic disorder. $T/W=0.001$.}
\label{fig:2}
\end{figure*}

To understand this, Fig. \ref{fig:2} compares the currents flowing through isotropic and anisotropic ($\xi_{\parallel}/\xi_{\bot}=6$) structures at $T/W=0.001$. Both structures have an identical energetic disorder. For the former, the current spans uniformly over the interior, and the conduction path acquires different distances, consistent with VRH theory.\cite{Shk_book} However, the anisotropic structure in Fig. 2(b) reveals nearest-neighbor inter-plane hopping along the transport direction. Conduction is dominated by a single path that consists of a chain of resistors connecting neighbor planes in a series. That path carries even more current (less branching) when compared to the isotropic structure.   

Reentrance to the activation regime at low $T$ for out-of-plane transport can be also understood from the percolation theory with the following argument. In the percolation theory,\cite{Amb71, Shk_book} a critical subnetwork is constructed from bonds (resistors) that satisfy the inequality
\begin{equation} \label{eq:perc}
 \frac{r_{ij}}{r_{max}}+\frac{|E_i|+|E_j|+|E_i-E_j|}{2E_{max}}<1,
\end{equation}
where
\begin{equation} \label{eq:perc2}
E_{max}=k_BT\ln(\frac{\gamma_0}{\Gamma_c})
\end{equation}
and
\begin{equation} \label{eq:perc25}
r_{max}^2=\frac{x^2+z^2}{r_{max\parallel}^2}+\frac{y^2}{r_{max\bot}^2}
\end{equation}
bounds an ellipsoid (oblate spheroid) with semi-major and semi-minor axises
\begin{align} \label{eq:perc3}
 r_{max\parallel}&=\frac{\xi_{\parallel}}{2}\ln\frac{\gamma_0}{\Gamma_c},\\
 r_{max\bot}&=\frac{\xi_{\bot}}{2}\ln\frac{\gamma_0}{\Gamma_c}. \label{eq:perc33}
\end{align}
$\Gamma_c$ is chosen such that the set of connected bonds is just enough for the subnetwork to span through the device, from the source to drain electrodes. This percolation criterion is satisfied at
\begin{equation} \label{eq:perc4}
 nr_{max\parallel}^2r_{max\bot}=v_c
\end{equation}
where $n=2\rho_0E_{max}$ is the total number of states per unit volume with $|E_i|<E_{max}$. $v_c$ is a dimensionless constant related to the critical density of the percolation problem. For a given site $i$, the factor $r_{max\parallel}^2r_{max\bot}$ allows all the states contained inside the ellipsoid centered at $i$ to create a bond. Note that the elliptical shape of $r_{max}$ results from the wave function anisotropy in \eqref{eq:3}. For the isotropic case, this ellipsoid transforms into a sphere of radius $r_{max}$, and the coordinate terms in \eqref{eq:perc4} are replaced by $r_{max}^3$.\cite{Amb71} If the localized states are strongly anisotropic $\xi_{\parallel}/\xi_{\bot}\gg1$ and positional disorder is weak $\Delta r<\xi_{\bot}$, the states in the $y$-direction, which fall inside the ellipsoid \eqref{eq:perc25} and are thus allowed to create a bond at the percolation threshold, 
belong to the nearest-neighbor $xz$-planes. This allows one to replace $r_{max\bot}$ in \eqref{eq:perc4} by the lattice constant $l$, which is the minimal bond length at percolation.
\begin{equation} \label{eq:perc5}
 \ln\left(\frac{\Gamma_c}{\gamma_0}\right)\approx\frac{v_c}{2\rho_0k_BT\xi_{\parallel}^2l}
\end{equation}
Since $y$ is the transport direction and the $xz$ tails of the wave functions from different planes do not overlap, $r_{max\parallel}\approx \xi_{\parallel}$. Within the $xz$ planes there are many strongly coupled states available to adjust the subnetwork such that a pair of states from the nearest neighbor planes with the smallest energy difference is to be  chosen to form a bond. For an electron traversing through the system this means that it is energetically favorable to hop in the $xz$ plane until the next vacant site on the other plane becomes closest in energy. From \eqref{eq:perc5}, an activated $T$ dependence of conductivity ($\sigma\propto \Gamma_c$) is obtained. 

\begin{figure}[h]
\includegraphics[keepaspectratio,width=0.9\columnwidth]{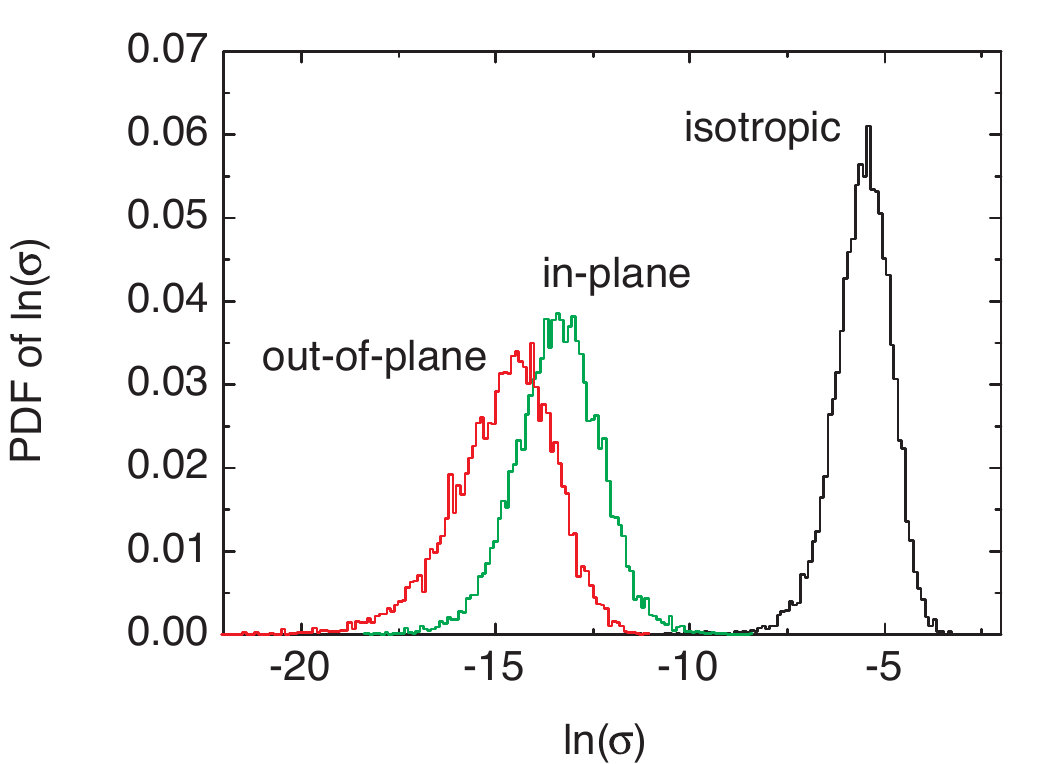}
\caption{(Color online) Probability distribution function of conductance fluctuations for $T/W=0.001$. $\xi_{\parallel}/\xi_{\bot}=6$ for in-plane and out-of-plane $\sigma$.}
\label{fig:3}
\end{figure}

Additional information on the conduction mechanism in the activation regime can be obtained from the probability distribution function (PDF) of the conductance fluctuations.\cite{Hug96} The hopping transport generally implies strong fluctuations as any external parameter (e.g., the chemical potential) varies because of an extremely broad distribution of elementary resistors composing the network.\cite{He03} In the activated regime, however, fluctuations are expected to be smaller than those in the VRH regime, since the bond length does not fluctuate. To check whether this holds for a low-$T$ activated regime, Fig. \ref{fig:3} shows PDF of the $\ln\sigma$ fluctuation for isotropic and anisotropic structures at $T/W=0.001$. In the activated (NNH) regime, $\sigma$ reveals strong fluctuations, comparable in magnitude with fluctuations in VRH regime. This might be understood to be a result of an additional constraint imposed by the wave function anisotropy (anisotropic breaks) on the current path, where this path has to adjust in a way shown in Fig. \ref{fig:2}(b). Note that the geometrical constraint due to reducing dimensionality generally enhances fluctuations, see Ref. \onlinecite{Hug96} and references therein, and leads ultimately to large non-self-averaging fluctuations in 1D.\cite{Rod09}

PDF is asymmetric and skewed to the right, which indicates that the samples with large $\sigma$ dominate the ensemble averaged $\sigma$. As $N\rightarrow\infty$, fluctuations decrease (not shown) and become negligible compared to the average value; PDF approaches a Gaussian distribution, in agreement with the central limit theorem. For isotropic and in-plane transport, PDF is already closely approximated by a Gaussian, which indicates that $N$ chosen is sufficiently large.

Relevant results were obtained by Nardes \textit{et al}.\cite{Nar07_PRB, Nar07_AM} in experiments on anisotropic PEDOT films where co-existing activated and VRH transport regimes were found. Their samples were prepared by spin coating and confirmed by scanning tunnel microscopy to contain elongated PEDOT grains aligned in horizontal layers and separated by poly(4-styrenesulfonate) (PSS) lamellas.\cite{Nar07_AM, Nar08} PEDOT grains possess good electrical conduction while PSS acts as an insulating barrier.\cite{Kim12, Nar07_PRB, Nar07_AM, Nar08, Cri06} Experimentally\cite{Nar07_PRB} extracted in-plane $T_0=3.2\times10^5$ K exceeds out-of-plane $T_0=70$ K, which is consistent with the result obtained above. Additional non-Ohmic measurements revealed the characteristic hopping length $\approx 1$ nm for the out-of-plane direction. This agrees with the plane-to-plane separation of PEDOT layers obtained for relaxed geometries in the first-principles calculations.\cite{Len11} Thus, activated out-of-plane conduction and low values of $\sigma_{\bot}$ ($\sim 10^{-6}$ S/cm) in experiment might be related to a strong charge localization and short-range order in the PEDOT layer across the thin film.\cite{Nar07_PRB, Nar07_AM} In-plane VRH in the measurement of the same sample, along with a larger $\sigma_{\parallel}$ ($\sim10^{-4}$ S/cm), might be explained by weaker localization, where the wave function extends along the polymer backbone and couples strongly with another state in a neighbor polymer unit. Note that, in experiment, $\alpha=1/4$ indicating 3D VRH, while the above theoretical results predict 2D. This might be attributed to the fact that  for in-plane electrical measurements the electrodes were placed 1 mm apart from each other, thus including many ($\sim10^6$) localized states composing a conductive network that is unlikely to maintain long-range order, in contrast to theoretical results where the long-range order (no positional disorder) was realized. A quantitative agreement with experiment\cite{Nar07_PRB, Nar07_AM} might be achieved for other parameters: $k_BW$ = 0.25-1.25 eV, which is of the order of the band gap of pristine PEDOT\cite{Len11}; $\gamma_0=10^{13}$ $s^{-1}$ --- a typical value for organic semiconductors\cite{Kim12_Pipe}; $l=1$ nm.

Finally, several comments are as follows. First, the results presented above were obtained for constant DOS, which might be a poor approximation for DOS in real polymeric systems.\cite{dos} Eq. \eqref{eq:1} was derived while assuming that transport occurs in a narrow energy band where DOS can be regarded as a constant for sufficiently low $T$.\cite{Mot69, Shk_book} For sufficiently low $T$, \eqref{eq:1} is still expected to hold true, even for DOS of strongly varying Gaussian shape.\cite{Zvy08} Because an overwhelming amount of experiments support Mott's law \eqref{eq:1}, the above results are expected to stay qualitatively the same also for different DOS shapes fulfilled with a low $T$ condition. Second, if positional disorder is added to the modeling with a deviation of 80\% relative to $l$,\cite{Jac11} the activated regime disappears, consistent with the traditional VRH theory.\cite{Shk_book} In this case of strong positional disorder, charge carriers propagate zig-zag like through the network. Third, to reproduce the absolute values of $\sigma$ in Fig. \ref{fig:2}, with arbitrary units converting to S/cm, $\gamma_0=10^{13}$ $s^{-1}$ should be used. Fourth, the above theory does not include a Coulomb interaction that is known\cite{Efr75} to create a soft gap in DOS near the Fermi energy and make $\alpha=1/2$ in \eqref{eq:1}. Electron interactions are expected to become important at low $T$, below the range where VRH occurs, and also if screening is not strong. This effect might be a topic of a separate study.  Fifth, the hopping rates \eqref{eq:3} assume electrons or holes as charge carriers. These rates are modified when polaron effects become important,\cite{Mar56} which also deserves a separate study.

In conclusion, numerical calculations of hopping conduction have shown that both activated temperature dependence and stretched exponential dependence of the the Mott's law \eqref{eq:1} should be observable in anisotropic structures at low temperatures. This implies nearest-neighbor and variable-range hopping for different transport directions. Both are characterized by conductance fluctuations of comparable amplitudes. Activated behavior (nearest-neighbor hopping) is a result of a single conduction path formation that adjusts in the planes where the wave functions strongly overlap. This has been demonstrated by current path visualization and using the percolation theory. These findings provide a microscopic explanation of anisotropic hopping conduction in PEDOT thin films observed by Nardes et al.\cite{Nar07_PRB, Nar07_AM}

\section{Acknowledgement}

This work was supported by the Energimyndigheten and NSC (SNIC 2015/4-20). It is a pleasure to acknowledge discussions with M. Kemerink.

\end{document}